# Quantum Text Classifier - A Synchronistic Approach Towards Classical and Quantum Machine Learning

Dr. Prabhat Santi
p.santi@tcs.com
*Tata Consultancy Services*

Kamakhya Mishra
kamakhya.mishra@tcs.com
*Tata Consultancy Services*

Sibabrata Mohanty
s.mohanty@tcs.com
*Tata Consultancy Services*

*Abstract*—Although it will be a while before a practical quantum computer is available, there is no need to hold off. Methods and algorithms are being developed to demonstrate the feasibility of running machine learning (ML) pipelines in QC (Quantum Computing). There is a lot of ongoing work on general QML (Quantum Machine Learning) algorithms and applications. However, a working model or pipeline for a text classifier using quantum algorithms isn't available. This paper introduces quantum machine learning w.r.t text classification to readers of classical machine learning. It begins with a brief description of quantum computing and basic quantum algorithms, with an emphasis on building text classification pipelines.

A new approach is introduced to implement an end-to-end text classification framework (Quantum Text Classifier - QTC), where pre- and post-processing of data is performed on a classical computer, and text classification is performed using the QML algorithm. This paper also presents an implementation of the QTC framework and available quantum ML algorithms for text classification using the IBM Qiskit library and IBM backends.

## I. INTRODUCTION

Traditional hardware has grown increasingly infeasible as scaling rules (Moore's law, Dennard scaling, and so on) have gotten saturated or have reached their limits. This has compelled computer firms to refocus on quantum hardware. Gradual improvements in nanotechnology and manufacturing, along with widely recognized potential benefits of quantum computing, have sparked significant interest and investment from the computer industry and allied communities.

Even though a full-fledged quantum computer is several years away, the availability of Noisy Intermediate-Scale Quantum (NISQ) [1] computers has provided the ability to early adopters to try out possibilities in quantum computing. Despite having a limited qubit count and being plagued by noise, NISQ computers are already capable of outperforming classical computers on specialized sampling jobs. [2] [3] [4]. With rapid advances being made, a practical error-mitigated NISQ computer is only a few years ahead.

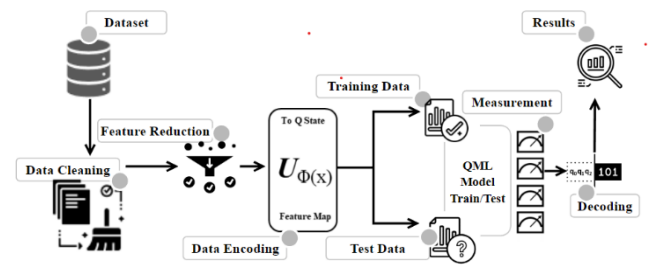

*Figure 1 – An end-to-end text classifier framework (Quantum Text Classifier – QTC) demonstrating the co-existence of classical and quantum computing platforms.*

The intersection of machine learning and quantum computing is one of the promising applications of quantum computing. In this paper, we present a hybrid framework leveraging quantum and classical ML algorithms for text classification. It begins by giving a quick overview of quantum computing and the relevant quantum algorithms with an emphasis on building a pipeline for text classifiers. Then, it introduces an end-to-end text classifier framework (Quantum Text Classifier – QTC) and demonstrates the implementation of the same using available quantum ML algorithms from the IBM Qiskit library *[Figure 1]*. The paper provides a framework for classical ML engineers to implement text classification solutions using QML models.

## II. QUANTUM MACHINE LEARNING

The intersection of Quantum computing and machine learning leads to four approaches based on the combination of classical (C) or Quantum (Q) data and classical (C) or Quantum (Q) processing [5] [6]. This paper focuses on the CQ scenario, which deals with classical data and quantum machine learning algorithms.

We highlight the usage of different quantum and classical machine learning techniques with a focus on text classification. Consequently, we provide a novel framework to use Quantum Classifier models on classically preprocessed data.

### A. *Classical Text Classification Algorithms*

In real-world text classification problems, supervised learning is more prevalent than unsupervised learning. The algorithm in supervised learning uses the attribute called features to classify the object. As part of the experiment, we used the major supervised learning algorithms from classical ML to set a benchmark for our hybrid quantum-classical text classification approach [6] [7].

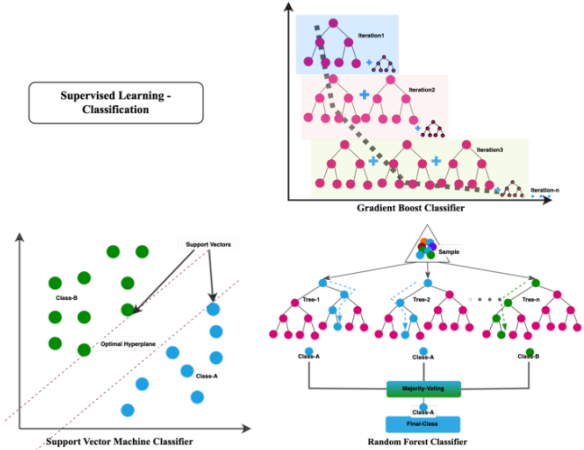

*Figure 2 - The three major classical supervised classification algorithms (Support Vector Classifier (SVC), Random Forrest Classifier (RFC) and Gradient Boost Classifier (GBC)) are used to set a benchmark for the Quantum Text Classifier framework proposed in this paper.*

### B. Quantum Text Classification Algorithms

In this section, we define the various quantum classifier models which can be effectively used in the QTC framework.

#### i) Quantum Support Vector Classifier (QSVC)

Support Vector Classifiers (SVC) are basically meant for binary classification problems. But the same principle can be utilized for multiclass classification after breaking down the multi-classification problem into smaller sub-problems, all of which are binary classification problems.

The QSVC is also known as a quantum-enhanced technique [8] because the quantum algorithm uses a feature map to draw the data points to quantum circuits executed by a quantum processor. However, QSVC works in the same way as SVC does.

#### ii) Variational Quantum Classifier (VQC)

The Variational Quantum Classifier (VQC) is a key supervised QML algorithm widely used for classification problems on a NISQ device. This model is particularly helpful on NISQ devices because it allows implementation without the need for additional error-correction approaches [9]. A VQC implementation consists of the encoding feature map circuit, a variational circuit (or parameterized circuit) called the Ansatz, and the measurement component. Iterative device measurements are used to calculate the cost function. Classical input data is mapped to a quantum feature space, which is based on quantum circuits (feature maps).

#### iii) Quantum Neural Network Classifier (QNNC)

Quantum neural networks are a subclass of variational quantum algorithms comprising quantum circuits that contain parameterized gate operations [10]. Feature map circuits map classical input data to a quantum feature space.

Several interesting quantum algorithms for supervised learning [11] have not been covered in this study. Nevertheless, the framework and ideas elucidated in this study will help the reader implement the other quantum algorithms that are not included here.

### III. THE QUANTUM TEXT CLASSIFIER (QTC) FRAMEWORK

In this section, we demonstrate the Quantum Text Classifier framework, a composite approach to training and inferencing machine learning models that use both classical and quantum computers. This framework depicts high-level abstractions of various building blocks for designing and training machine learning models on classical and/or quantum computers, as appropriate.

We leveraged a classical computer for pre-processing, feature extraction and feature reduction, and post-processing for inferencing the model outcome in our implementations. A quantum computer or simulator was used to encode the training data, create the custom feature map, train the quantum model, mitigate errors, optimize, and post-process the data.

Proving the exponential benefits of using Quantum Machine Learning Models Vs. Classical Models is beyond the scope of this study.

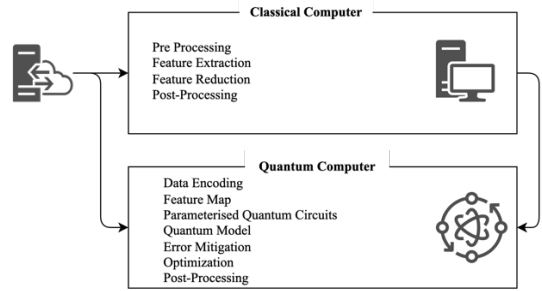

*Figure 3 - High Level QTC hybrid framework proposed in this paper.*

### A. Pre-Processing and Data preparation

The text data is preprocessed and cleaned to remove any irrelevant information.

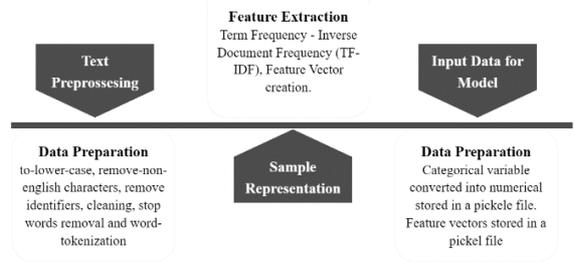

*Figure 4 - QTC Pre-Processing and Input Data Creation.*

Firstly, the categorical variables are converted into numerical forms (label-encoding). Secondly, the preprocessed text input is converted into a matrix of TF-IDF (term frequency-inverse document frequency) features [12]. The TF-IDF algorithm is a simple approach of weighting words (or phrases, lemmas, etc.) by using the frequency of the words to identify their relevance in a document amongst a corpus (collection of documents). The resulting numerical feature vector can now be used by the quantum algorithm as input.

The outputs of the step are saved as two pickle files to avoid running the preprocessing step for every ML iteration.

### B. Feature Reduction

The number of features is reduced to the number of error-free qubits we are working with. Since our selected categories or labels are less in count and the feature file now doesn't have the labels, the Principal Component Analysis (PCA) technique was used for feature selection. The intent of this process is to reduce the number of features or dimensions and identify a smaller set of features that can still describe the variations in the data [13].

### C. Data Encoding Circuit - Feature Map

Each classical feature vector $x$ from the sample input from the previous step is now mapped to a quantum state $|\Phi(x)\rangle\langle\Phi(x)|$ nonlinearly using a quantum feature map. This quantum states or feature vectors are vectors in a Hilbert space [11].

There are multiple algorithms or techniques available which can be used to map the input data into a quantum state. The most popular is the quantum frequency encoding technique, where each feature vector is associated with a specific quantum state based on its frequency.

As part of the QTC framework, we propose the use of the 1$^{st}$ order (ZFeatureMap) and 2$^{nd}$ order (ZZFeatureMap) representation of the Pauli expansion circuit as prescribed by the authors of [9]. These feature map circuits are difficult to simulate using classical computers but at the same time as they have short depth, can be successfully implemented on available NISQ devices.

Here the quantum feature map on $n$-qubits of depth $d$ is implemented by the unitary operator
$U\Phi(x) = \prod_d U\Phi(x) H^{\otimes n}$, where H denotes the Hadamard gate and $U\Phi(x)$ is a diagonal unitary gate in the Pauli Z basis
$U\Phi(x) = exp\big(i \sum_{S\subseteq[n]} \Phi S(x) \prod_{k\in S} P_k\big)$

The implementation of ZFeatureMap and ZZFeatureMap circuits for two qubits in IBM Qiskit is shown in *Figure 5* below.

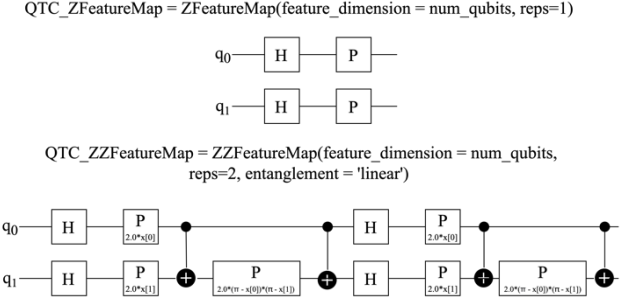

*Figure 5 - ZFeaturemap and ZZFeaturemap circuit implementation for two qubits in IBM Qiskit.*

### D. Quantum Classification

Based on the quantum states created in step [III. C. ], one of the quantum classifier algorithms explained in [II. B. ] is utilized to classify the input data.

The objective of the quantum support vector classifiers is to find a hyperplane separating each vector $|\varphi(xi)\rangle\langle\varphi(xi)|$ depending on its label and supported by a reduced number of support vectors [11] [12].

In the case of VQC, a variational circuit (i.e., parameterized circuit) or Ansatz is used, whose parameters can be altered by a classical optimizer after computing the VQC's cost function or objective function. The process iterates with the changed parameters until the termination condition is met [9].

In the case of Quantum Neural Network Classifiers [10], the states created by the feature map are further evolved via a variational form $|g\theta(x)\rangle := G\theta|\Phi(x)\rangle$, where the parameters $\theta \in \Theta$ are selected to minimize a certain loss function [12, 2].

### E. Post Processing – Decoding and Labelling

The quantum states are measured, and the findings are utilized to predict the category or class of the text data. The findings are decoded to arrive at a final prediction about the class of the text data. The quantum classifier's predictions are compared to the true labels of the input data to evaluate the classifier's accuracy.

## IV. IMPLEMENTATION AND RESULTS

In this section, we present an implementation of the QTC framework using the classical and quantum ML models briefed in section II. The environment setup for the experiments is detailed in the table in Appendix VIII. (*Figure 13*).

### A. Input Dataset

The dataset used in this experiment aims to emulate a real business case scenario for text classification. It contains actual resumes (Column - Resume_str) and the category (Column – Category) of the corresponding candidate. The data has been sanitized to remove any personal identifiers. The ML models are expected to identify the category based on the features generated from the resume text. The resume text in column Resume_str is used in the preprocessing step of the ML pipeline to generate the features.

A snapshot of the input dataset is presented in *Figure 6*.

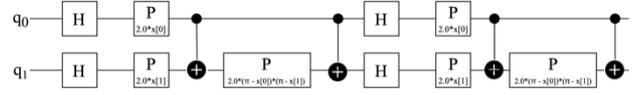

*Figure 6 – Input dataset used in the experiment.*

### B. Preprocessing and Feature Reduction

The input dataset is analyzed, and samples from three top categories having equal distribution are selected. We follow the steps outlined in sections III. A. and III. B. to first preprocess the input samples and then generate the numerical TF-IDF matrix of feature vectors. The module feature_extraction.text.TfidfVectorizer from the scikit-learn (or sklearn) library [12] is used to convert the samples of text data into a matrix of TF-IDF features. A maximum of 20 features are retained in the matrix. We also converted the categorical variables into numerical ones. As a result, two intermediate pickle files are generated with vectorized numerical features and numerical categories, respectively. These files would be used as inputs for the implementation of the models in the subsequent sections of the QTC pipeline.

As prescribed in section III. B. the PCA technique is used to further reduce the input features to two, which is equal to the number of qubits we would be using to execute the quantum algorithms.

The input feature samples and labels are split into two sets for training and testing the models in the next steps. The ratio of the split is kept at 20 % for the test set and 80 % for the train set.

### C. Classical Model Implementation

The train and test files generated in section B. are used to train the three classical models proposed in section II. A. The three classical algorithms - Support Vector Classifier with polynomial kernel of degree 3, Gradient Boosting Classifier and Random Forest Classifier - are implemented via multiple iterations of the QTC framework so that the results can be used as a benchmark. The results are listed for comparison in *Figure 12*.

### D. Quantum Model Implementation

The train and test files generated in section B. are used as inputs in multiple iterations of the QTC framework to train the three quantum models proposed in section II. B.

The input feature samples from the previous section need to be mapped to quantum states in a Hilbert space using a Feature Map as explained in sectionC. III. C. In our implementations, the inbuilt feature map ZZFeatureMap from Qiskit as shown in *Figure 7,* is used as the data encoding circuit.

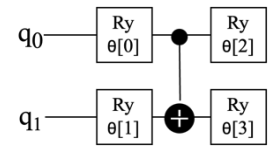

*Figure 7 - The ZZFeatureMap circuit with 2 qubits, 2 repetition and linear entanglement which is used for the QSVC kernel.*

The encoded data is then used to train the three Quantum Models using IBMQ simulator "qasm_simulator" as described in the following subsections.

*i)   Quantum Support Vector Classifier (QSVC)*

To implement QSVC, a kernel for the Support Vector Classifier is built out of the quantum states generated by the quantum feature map.

The kernel is used to train and test the QSVC model using the IBMQ simulator. The same optimized kernel is used while classifying new unlabeled data.

*ii)   Variational Quantum Classifier (VQC)*

To implement VQC, the RealAmplitudes circuit given in *Figure 8* is used as the Ansatz or the variational circuit. This Ansatz circuit consists of alternating layers of Y rotation and CX entanglements. In our case, we have used only one repetition.

The quantum states generated by the quantum feature map are fed into the Ansatz, and a classical optimizer analyzes the output. A Classical optimization process, COBYLA (Constrained Optimization by Linear Approximation Optimizer), was used to optimize the values of the variational circuit. The whole process is repeated with the changed parameters until the cost function value decreases.

*Figure 8 - RealAmplitudes Circuit with 1 layer or repetition of Y rotation and CX entanglement for 2 qubits used as Ansatz for VQC implementation.*

The learning curve for the VQC training process with 30 iterations is shown in *Figure 9.*

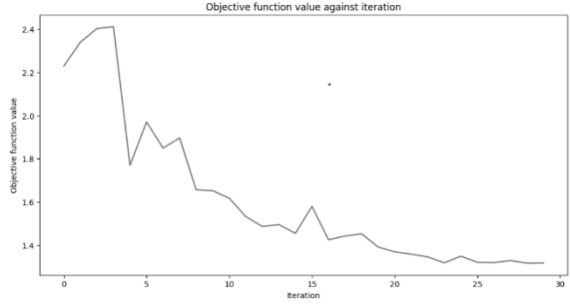

*Figure 9- The learning curve of the VQC algorithm for 30 iterations.*

*iii) Quantum Neural Network Classifier (QNNC)*

To implement QNNC, the RealAmplitudes circuit in *Figure 8* is used as the Ansatz, i.e., the variational circuit (like the usage in VQC). The artificial quantum neural network is constructed using the composite Featuremap and Ansatz circuit, as shown in *Figure 10* and the corresponding parameters. The Qiskit Neural Network Classifier algorithm is then used to train and use this quantum neural network. Classical optimization process COBYLA (Constrained Optimization by Linear Approximation optimizer) was used to optimize the parameters of the Ansatz.

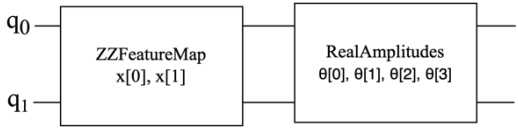

*Figure 10 - The composite Feature map and Ansatz quantum circuit used for the QNNC implementation*

The learning curve for the QNNC training process with 30 iterations is shown in *Figure 11,* and the comparative performance report is shown in *Figure 12*.

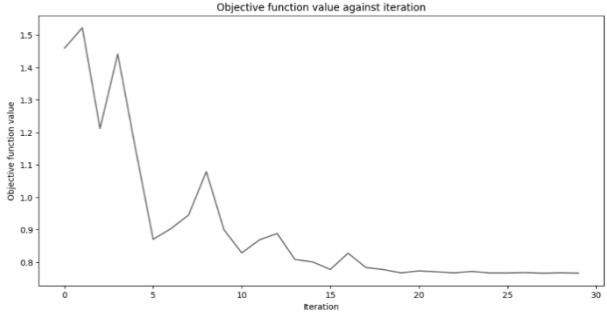

*Figure 11 - The learning curve of the QNNC algorithm for 30 iterations.*

*E. Results and Inferences*

Classical and quantum machines are used to run the models and validate the proposed QTC framework, as described in section III.

The results and inferences from the iterations and experiments using the QTC framework, as detailed in section C. and D. are listed below in *Figure 12* for comparison.

| Evaluation Matrix | Classical Classifier Algorithm | | | Quantum Classifier Algorithm | | |
|---|---|---|---|---|---|---|
| | Support Vector | Random Forest | Gradient Boosting | Quantum Support Vector | Variational Quantum Classifier | Quantum Neural Network Classifier |
| accuracy | 0.903 | 0.861 | 0.847 | 0.889 | 0.806 | 0.569 |
| precision | 0.917 | 0.866 | 0.856 | 0.892 | 0.831 | 0.694 |
| recall | 0.903 | 0.861 | 0.847 | 0.889 | 0.806 | 0.569 |
| f1-score | 0.904 | 0.86 | 0.848 | 0.89 | 0.805 | 0.531 |

*Figure 12 - The results are listed for comparison from the implementation of the three classical and three Quantum models using the QTC framework.*

The pipeline will support the extraction of feature from any text data with minimal changes to the code in the preprocessing step detailed in section B. This framework can easily accommodate any new classical or quantum algorithms to validate the same dataset.

## V. CONCLUSION AND FUTURE WORK

In this paper, we presented an approach to building a hybrid – classical and quantum text classifier (QTC) pipeline, which should be easy to follow and implement by existing classical machine learning engineers and architects.

This framework supports the extraction of features from any text data with minimal change in the code used in section III. A.

Popular Quantum Natural Language Processing (QNLP) techniques, which are based on ZX-calculus [16], have the limitation of being applicable only to toy-sized datasets for the time being. But the QTC framework highlighted in this paper already attempts the classification of larger documents like resumes using a hybrid classical and quantum-enhanced framework.

We have also identified possibilities for improvement in the individual stages of the framework.

*A. Next Steps*

The work as part of this paper provides the motivation to investigate the following:
1. The current QTC framework is optimized for three categories and two features [III. C. ] due to the limitations of available NISQ Quantum computers. Further research needs to be done to increase the number of categories and features.
2. The use of LDA (Linear Discriminant Analysis) over PCA must be investigated to identify the best method for feature selection. [III. B. ]
3. Quantum natural language processing techniques, such as quantum embedding, should be investigated to enhance the performance of the quantum blocks in the pipeline.
4. New ways need to be identified and experimented with to train quantum models with larger datasets or find ways to improve accuracy with smaller datasets.
5. Later, with enough resources, the ZX-calculus-based approach needs to be investigated and considered as a candidate for the quantum classifier within the framework.
6. Custom Feature Maps should be identified and tested in step [III. C. ] to improve the feature encoding process.
7. The use of Parameterized Feature Maps should be investigated in step [III. C. ] to provide flexibility to the feature encoding stage.
8. Different quantum machines should be explored to identify improvement in accuracy with scaling.

## VI. ACKNOWLEDGMENT

The authors gratefully acknowledge the reviews and proof reading by Sayantan Pramanik (Researcher) and Girish Chandra (Principal Scientist) from TCS Research and Incubation Unit, which has helped the paper reach its final state.

# VIII. APPENDIX

The dataset used and the code is available on request from the authors.

The complete environment details for the QTC framework are given in table below.

| Qiskit Software | Version |
|---:|---:|
| qiskit-terra | 0.22.0 |
| qiskit-aer | 0.11.0 |
| qiskit-ibmq-provider | 0.19.2 |
| qiskit | 0.39.0 |
| qiskit-nature | 0.4.2 |
| qiskit-finance | 0.3.3 |
| qiskit-optimization | 0.4.0 |
| qiskit-machine-learning | 0.4.0 |
| **System information** | |
| Python version | 3.8.13 |
| Python compiler | GCC 10.3.0 |
| Python build | default, Mar 25 2022 06:04:10 |
| OS | Linux |
| CPUs | 8 |
| Memory (Gb) | 31.211315155029297 |
| | Tue Nov 01 12:32:45 2022 UTC |

*Figure 13 - The environment details for the QTC framework used for the experiments.*

The complete results and score for the experiments is given in the below table.

```
                          precision  recall  f1-score  support
df_CML01_SVC  0              0.793   0.958     0.868   24.000
              1              0.957   0.917     0.936   24.000
              2              1.000   0.833     0.909   24.000
              accuracy       0.903   0.903     0.903    0.903
              macro avg      0.917   0.903     0.904   72.000
              weighted avg   0.917   0.903     0.904   72.000
df_CML02_RFC  0              0.778   0.875     0.824   24.000
              1              0.920   0.958     0.939   24.000
              2              0.900   0.750     0.818   24.000
              accuracy       0.861   0.861     0.861    0.861
              macro avg      0.866   0.861     0.860   72.000
              weighted avg   0.866   0.861     0.860   72.000
df_CML03_GBC  0              0.750   0.875     0.808   24.000
              1              0.917   0.917     0.917   24.000
              2              0.900   0.750     0.818   24.000
              accuracy       0.847   0.847     0.847    0.847
              macro avg      0.856   0.847     0.848   72.000
              weighted avg   0.856   0.847     0.848   72.000
df_QML01_QSVC 0              0.808   0.875     0.840   24.000
              1              0.957   0.917     0.936   24.000
              2              0.913   0.875     0.894   24.000
              accuracy       0.889   0.889     0.889    0.889
              macro avg      0.892   0.889     0.890   72.000
              weighted avg   0.892   0.889     0.890   72.000
df_QML02_QNN  0              0.800   0.167     0.276   24.000
              1              0.449   0.917     0.603   24.000
              2              0.833   0.625     0.714   24.000
              accuracy       0.569   0.569     0.569    0.569
              macro avg      0.694   0.569     0.531   72.000
              weighted avg   0.694   0.569     0.531   72.000
df_QML03_VQC  0              0.697   0.958     0.807   24.000
              1              0.895   0.708     0.791   24.000
              2              0.900   0.750     0.818   24.000
              accuracy       0.806   0.806     0.806    0.806
              macro avg      0.831   0.806     0.805   72.000
              weighted avg   0.831   0.806     0.805   72.000
```

*Figure 14 - The results for classical algorithms are labeled as df_CML**(Classical Machine Learning) and quantum algorithms are labeled as df_QML**(Quantum Machine Learning)*